\documentclass[english,conference,compsocconf]{IEEEtran}
\usepackage[T1]{fontenc}
\usepackage[latin9]{inputenc}
\usepackage{float}
\usepackage{amstext}
\usepackage{graphicx}
\usepackage{setspace}

\makeatletter

\floatstyle{ruled}
\newfloat{algorithm}{tbp}{loa}
\providecommand{\algorithmname}{Algorithm}
\floatname{algorithm}{\protect\algorithmname}

\IEEEoverridecommandlockouts

\usepackage{subfigure} 

\makeatother

\usepackage{babel}
\begin{document}

\title{Opportunistic Routing Based on Daily Routines}

\author{\IEEEauthorblockN{Waldir Moreira, and Paulo Mendes} \IEEEauthorblockA{SITI,
University Lusófona, Portugal\\
 \{waldir.junior, paulo.mendes\}@ulusofona.pt} \and\IEEEauthorblockN{Susana
Sargento} \IEEEauthorblockA{IT, University of Aveiro\\
susana@ua.pt}\thanks{''This is the author's preprint version. Personal
use of this material is permitted. However, permission to reprint/republish
this material for advertising or promotion or for creating new collective
works for resale or for redistribution to thirds must be obtained
from the copyright owner. The camera-ready version of this work has
been published at AOC 2012, date of June 2012 and is property of IEEE.''} }
\maketitle
\begin{abstract}
Opportunistic routing is being investigated to enable the proliferation
of low-cost wireless applications. A recent trend is looking at social
structures, inferred from the social nature of human mobility, to
bring messages close to a destination. To have a better picture of
social structures, social-based opportunistic routing solutions should
consider the dynamism of users' behavior resulting from their daily
routines. We address this challenge by presenting \emph{dLife, }a
routing algorithm able to capture the dynamics of the network represented
by time-evolving social ties between pair of nodes. Experimental results
based on synthetic mobility models and real human traces show that
\emph{dLife} has better delivery probability, latency, and cost than
proposals based on social structures.\end{abstract}


\section{Introduction}

The pervasive deployment of wireless personal devices is creating
the opportunity for the development of novel applications. The exploitation
of such applications with a good performance-cost tradeoff is possible
by allowing devices to use free spectrum to exchange data whenever
they are within wireless range. Since every contact is an opportunity
to forward data, there is the need to develop routing algorithms able
to bring messages close to a destination, with high probability, low
delay and costs. Most of the proposed routing solutions focus on inter-contact
times alone \cite{impactHuman}, while there is still significant
investigation to understand the nature of such statistics (e.g., power-law,
behavior dependent on node context). Moreover, the major drawback
of such approaches is the instability of the created proximity graphs
\cite{bubble2011}, which changes with users' mobility.

A recent trend is investigating the impact that more stable social
structures (inferred from the social nature of human mobility) have
on opportunistic routing \cite{bubble2011,simbet}. Such social structures
are created based on social similarity metrics that allow the identification
of the centrality that nodes have in a cluster/community. This allows
forwarders to use the identified hub nodes to increase the probability
of delivering messages inside (local centrality) or outside (global
centrality) a community, based on the assumption that the probability
of nodes to meet each other is proportional to the strength of their
social connection.

A major limitation of approaches that identify social structures,
such as communities, is the lack of consideration about the dynamics
of networks, which refers to the evolving structure of the network
itself, the making and braking of network ties: over a day a user
meets different people at every moment. Thus, the user's personal
network changes, and so does the global structure of the social network
to which he/she belongs.

When considering dynamic social similarity, it is imperative to accurately
represent the actual daily interaction among users: it has been shown
\cite{thyneighbor} that social interactions extracted from proximity
graphs must be mapped into a cleaner social connectivity representation
(i.e., comprising only stable social contacts) to improve forwarding.
This motivates us to investigate a routing solution able to capture
network dynamics, represented by users' daily life routine. We focus
on the representation of daily routines, since routines can be used
to identify future interaction among users sharing similar movement
patterns, interests, and communities \cite{behaviour}. Existing proposals
\cite{socialcast,bubble2011,simbet} succeed in identifying similarities
(e.g., interests) among users, but their performance is affected as
dynamism derived from users' daily routines is not considered.

To address this challenge, we propose \emph{dLife} that uses time-evolving
social structures to reflect the different behavior that users have
in different daily periods of time: \emph{dLife} represents the dynamics
of social structures as a weighted contact graph, where the weights
(i.e., social strengths) express how long a pair of nodes is in contact
over different period of times. It considers two complementary utility
functions: \emph{Time-Evolving Contact Duration} (TECD) that captures
the evolution of social interaction among pairs of users in the same
daily period of time, over consecutive days; and \emph{TECD Importance
}(TECDi) that captures the evolution of user's importance, based on
its node degree and the social strength towards its neighbors, in
different periods of time.

In this paper, we show the performance gain of \emph{dLife} against
proposals that are only aware of social structures and node centrality
metrics, e.g., \emph{Bubble Rap} \cite{bubble2011}. We also analyze
the impact that centrality metrics have on routing, since by determining
the relative importance of a node within the community such metrics
create potential bottleneck points. For that, we created a community-based
version of \emph{dLife}, called \emph{dLifeComm,} for a fair comparison
with \emph{Bubble Rap}. Results show that both versions of \emph{dLife}
manage to capture the dynamism of social daily behavior along with
social interaction strength, resulting in improved delivery probability,
cost, and latency. Our findings also highlight the impact that centrality
has on routing performance when comparing the performance of two community-based
approaches (\emph{dLifeComm} and \emph{Bubble} \emph{Rap}).

This paper is structured as follows. Section 2 briefly analyses the
related work. Section 3 presents \emph{TECD} and \emph{TECDi} utility
functions along with the algorithms for both versions of\emph{ dLife}.
Section 4 presents the evaluation methodology, setup, and results.
In Section 5 the paper is concluded and future work is presented.

\section{Related Work\label{sec:Related-Work}}

Most of the existing opportunistic routing solutions are based on
some level of replication \cite{latincom}. Among these proposals,
emerge solutions based on different representations of social similarity:
i) labeling users according to their social groups (e.g., \emph{Label}
\cite{label}); ii) looking at the importance (i.e., popularity) of
nodes (e.g., \emph{PeopleRank} \cite{people}); iii) combining the
notion of community and centrality (e.g., \emph{SimBet} \cite{simbet}
and \emph{Bubble} \emph{Rap} \cite{bubble2011}); iv) considering
interests that users have in common (e.g., \emph{SocialCast} \cite{socialcast}).

Such prior-art shows that social-based solutions are more stable than
those which only consider node mobility. However, they do not consider
the dynamism of users' behavior (i.e., social daily routines) and
use centrality metrics, which may create bottlenecks in the network.
Moreover, such approaches assume that communities remain static after
creation, which is not a realistic assumption. 

On the other hand, prior-art also shows that users have routines that
can be used to derive future behavior \cite{behaviour}. It has been
proven that mapping real social interactions to a clean (i.e., more
stable) connectivity representation is rather useful to improve delivery
\cite{thyneighbor}. With \emph{dLife}, users' daily routines are
considered to quantify the time-evolving strength of social interactions
and so to foresee more accurately future social contacts than with
proximity graphs inferred directly from inter-contact times.

\begin{figure*}
\begin{centering}
\includegraphics[scale=0.35]{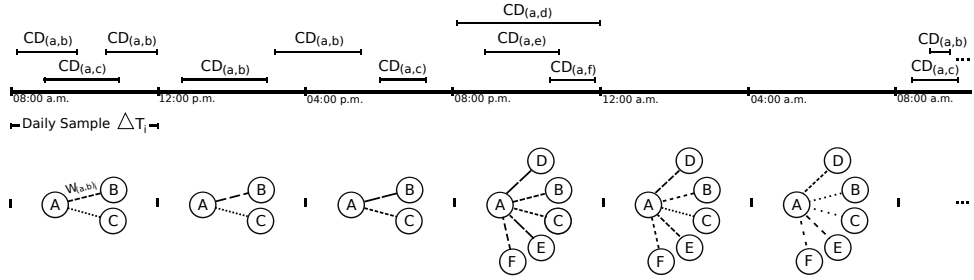}\vspace{-0.3 cm}
\par\end{centering}

\protect\caption{\label{fig:1-1-1}Contacts a user $A$ has with a set of users $x$
($CD_{(a,x)}$) in different daily samples $\Delta T_{i}$. }
\vspace{-0.5 cm}
\end{figure*}

\section{The \emph{dLife} Algorithm\label{sec:Proposal}}

The major motivation to devise social-based opportunistic routing
has to do with the higher stability that social similarity has in
comparison to inter-contact times. However, the dynamism of users'
social behavior (extracted from daily routines) should be considered
in order to guarantee a more realistic representation. This major
aspect is missing from existing social-based routing solutions, such
as \emph{Bubble Rap}.

Thus, we propose \emph{dLife} that uses two novel utility functions:
\emph{TECD} to forward messages to nodes that have a stronger social
relationship with the destination than the current carrier; with \emph{TECD}
each node computes the average of its contact duration with other
nodes during the same set of daily time periods over consecutive days.
Our assumption is that contact duration can provide more reliable
information than contact history, or frequency when it comes to identifying
the strength of social relationships. The reason for considering different
daily time periods relates to the fact that users present different
behavior during their daily routines \cite{behaviour}. If the carrier
and encountered node have no social information towards the destination,
forwarding takes place based on a second utility function, \emph{TECDi},
where the encountered node gets a message if it has greater importance
than the carrier.

A second version of \emph{dLife}, \emph{dLifeComm}, is designed to
allow an easier comparison of solutions that are focused on the dynamics
of the network (i.e., \emph{dLife}, based on users' daily routine)
and solutions that are focused on the structure of network (i.e.,
\emph{Bubble} \emph{Rap}, based on node centrality). \emph{dLifeComm}
uses \emph{TECD} and \emph{TECDi} to exploit communities that arise
from social interaction. Communities are detected based on the K-Clique
algorithm \cite{kclique}, as occurs with \emph{Bubble Rap}: \emph{TECD}
is used to forward within a community based on the social strength
that the carrier and encountered nodes have towards the destination,
and not their centrality; \emph{TECDi} is used to forward data based
on users' importance outside a community.

We start this section by explaining how K-Clique is used to detect
social structures (i.e., communities) by \emph{Bubble} \emph{Rap}
and \emph{dLifeComm}. Then, we explain how to capture the dynamics
of the network by computing \emph{TECD}/\emph{TECDi.} Finally, we
show how to use \emph{TECD} and \emph{TECDi} to create the \emph{dLife}
and \emph{dLifeComm} algorithms.

\subsection{Usage of Social Structures \label{sub:K-Clique-Community-Detection}}

A social structure defined as a K-Clique community \cite{kclique}
is a union of all cliques (complete subgraphs of size $k$) that can
be reached from each other through a series of adjacent cliques, where
cliques are adjacent if they share $k-1$ nodes. 

Both \emph{Bubble} \emph{Rap} and \emph{dLifeComm} use the K-Clique
algorithm to detect the social structure in a proximity graph. The
main difference between them is that the former uses the detected
structure to compute the centrality of nodes within and outside communities,
lacking a representation of the different levels of social interaction
that users have over different daily periods of time. On the other
hand, \emph{dLifeComm} considers continuously updated social information,
computed based on \emph{TECD} and \emph{TECDi}, for forwarding over
the detected social structure. That is, the fixed communities detected
are the same as in \emph{Bubble} \emph{Rap}, but the links considered
for forwarding within and between communities change over time as
they represent different levels of social strength in different time
periods. This means that while \emph{Bubble} \emph{Rap} considers
a fixed social structure, \emph{dLifeComm} is aware of its dynamics:
the network is still a fixed collection of linked individuals, but
now users' daily routines influence the way links are used.

Contrary to \emph{Bubble Rap} and \emph{dLifeComm}, \emph{dLife} does
not use any social network analysis algorithm, such as K-Clique, to
detect a fixed global social structure: \emph{dLife} relies on \emph{TECD}
and \emph{TECDi} utility functions to capture the dynamics of the
network by identifying time-evolving social structures that reflect
the different interactions that users have over different daily periods
of time. With \emph{dLife} the static structure of traditional network
analysis can be thought of as different snapshots taken during specific
periods of time.

\subsection{Time-Evolving Contact Duration (TECD)\label{sub:TECD}}

\emph{TECD} aims to capture the evolution of social interactions in
the same daily period of time (hereafter called daily sample) over
consecutive days, by computing social strength based on the average
duration of contacts.

Fig. \ref{fig:1-1-1} shows how social interactions (from the point
of view of user $A$) varies during a day. For instance, it illustrates
a daily sample (8 pm - 12 am) over which the social strength of user
$A$ to users $D$, $E$, and $F$ is much stronger (less intermittent
line) than the strength to users $B$ and $C$. Fig. \ref{fig:1-1-1}
aims to show the dynamics of a social network over a one-day period,
where users' behavior in different daily samples lead to different
social structures.

As illustrated in Fig. \ref{fig:1-1-1}, users' social strength in
a given daily sample depends on the average contact duration that
they have in such time period: if user $x$ has $n$ contacts with
user $y$ in a daily sample $\Delta T_{i}$, having each contact $k$
a certain duration (\emph{Contact} \emph{Duration} - $CD_{(x,y)_{k}}$),
at the end of $\Delta T_{i}$ the \emph{Total Contact Time} ($TCT_{(x,y)_{i}}$)
between them is given by Eq. \ref{eq:1}.\vspace{-0.3 cm}

\begin{equation}
TCT_{(x,y)_{i}}=\sum_{k=1}^{n}CD_{(x,y)_{k}}\label{eq:1}
\end{equation}
\vspace{-0.3 cm}

The \emph{Total Contact Time} between users in the same daily sample
over consecutive days can be used to estimate the average duration
of their contacts for that specific daily sample\,\cite{behaviour}:
the average duration of contacts between users $x$ and $y$ during
a daily sample $\Delta T_{i}$ in a day $j$ ($AD_{(x,y)_{i}}^{j}$)
is given by a cumulative moving average of their $TCT$ in that daily
sample ($TCT_{(x,y)_{i}}^{j}$) and the average duration of their
contacts during the same daily sample $\Delta T_{i}$ in the previous
day (cf.\,Eq.\,\ref{eq:2}).\vspace{-0.3 cm}

\begin{equation}
AD_{(x,y)_{i}}^{j}=\frac{{TCT_{(x,y)_{i}}^{j}+(j-1)AD_{(x,y)_{i}}^{j-1}}}{j}\label{eq:2}
\end{equation}
\vspace{-0.3 cm}

The social strength between users in a specific daily sample may also
provide some insight about their social strength in consecutive $k$
samples in the same day, $\Delta T_{i+k}$. This is what we call \emph{Time
Transitive Property}. This property increases the probability of nodes
being capable of transmitting large data chunks, since transmission
can be resumed in the next daily sample with high probability. 

The \emph{TECD} utility function (cf. Eq. \ref{eq:3}) is able to
capture the social strength ($w_{(x,y)_{i}}$) between any pair of
users $x$ and $y$ in a daily sample $\Delta T_{i}$ based on the
\emph{Average Duration }($AD_{(x,y)_{i}}$) of contacts between them
in such daily sample and in consecutive $t-1$ samples, where $t$
represents the total number of daily samples. When $k>t$, the corresponding
$AD_{(x,y)}$ value refers to the daily sample $k-t$. In Eq. \ref{eq:3}
the time transitive property is given by the weight $\frac{t}{t\text{+k-i}}$,
where the highest weight is associated to the average contact duration
in the current daily sample, being it reduced in consecutive samples.\vspace{-0.3 cm}

\begin{equation}
TECD=w_{(x,y)_{i}}=\sum_{k=i}^{i+t-1}\frac{t}{t+k-i}AD_{(x,y)_{k}}\label{eq:3}
\end{equation}
\vspace{-0.3 cm}

\subsection{TECD Importance (TECDi)\label{sub:TECDi}}

\emph{TECDi} aims to capture the \emph{Importance} ($I_{x}^{i}$)
of any user $x$ in a daily sample $\Delta T_{i}$, based on its social
strength (\emph{TECD}) towards each user that belongs to its neighbor
set ($N_{x}$) in that time interval, in addition to the importance
of such neighbors.

\emph{TECDi} is based on the \emph{PeopleRank} function \cite{people}.
However, \emph{TECDi} considers the social strength between a user
and its neighbors encountered within a specific $\Delta T_{i}$, while
\emph{PeopleRank} computes the importance considering all neighbors
of $x$ at any time. It is worth mentioning that the dumping factor
($d$) in \emph{TECDi} has a similar meaning as in \emph{PeopleRank:}
to introduce some randomness while taking forwarding decisions.\vspace{-0.3 cm}

\begin{equation}
TECDi=I_{x}^{i}=(1-d)+d\sum_{y\epsilon\, N_{x}}w_{(x,y)_{i}}\frac{{I_{y}^{i}}}{N_{x}}\label{eq:4}
\end{equation}
\vspace{-0.3 cm}

\subsection{Distributed Algorithm}

As mentioned before, \emph{dLife} decides to replicate messages based
on \emph{TECD}/\emph{TECD}i. If the encountered node has better relationship
with the destination in the current daily sample, it receives messages'
copies. By having higher weight (i.e., high social relationship),
there is a much greater chance that the encountered node will meet
the destination in the future. If relationship to destination is unknown,
replication happens only if the encountered node has higher importance
than the carrier.

\emph{dLife}'s operation happens as follows (cf. Alg. \ref{alg:Dlife}):
when the $CurrentNode$ meets a \emph{$Node_{i}$} in a daily sample
$\Delta T_{k}$, it gets a list of all neighbors of \emph{$Node_{i}$}
in that daily sample and its weights towards them (\emph{\scriptsize{}$Node_{i}$}{\scriptsize{}.WeightsToAllneighbors}
computed based on Eq. \ref{eq:3}). Then, every $Message_{j}$ in
$CurrentNode$'s buffer is replicated to \emph{$Node_{i}$ }if the
latter's weight towards the destination ({\scriptsize{}getWeightTo($Destination_{j}$)})
is greater than $CurrentNode$'s weight towards the same destination.
Otherwise, $CurrentNode$ receives \emph{$Node_{i}$}'s importance,
and messages are replicated if \emph{$Node_{i}$} is more important
than the $CurrentNode$ in the current $\Delta T_{k}$.\vspace{-0.2 cm}

\begin{algorithm}[H]
\protect\caption{\label{alg:Dlife}Forwarding with \emph{dLife}}

\textbf{\scriptsize{}begin}{\scriptsize \par}

{\scriptsize{}\hspace{0.1 cm}}\textbf{\scriptsize{}foreach}\emph{\scriptsize{}
$Node_{i}$ }{\scriptsize{}encountered by $CurrentNode$ }\textbf{\scriptsize{}do}{\scriptsize \par}

{\scriptsize{}\hspace{0.3 cm}receive(}\emph{\scriptsize{}$Node_{i}$}{\scriptsize{}.WeightsToAllneighbors)}\textbf{\scriptsize{}\vspace{-0.01 cm}}{\scriptsize \par}

{\scriptsize{}\hspace{0.3 cm}}\textbf{\scriptsize{}foreach}{\scriptsize{}
$Message_{j}$}\textbf{\scriptsize{} }{\scriptsize{}$\in$}\textbf{\scriptsize{}
}{\scriptsize{}buffer.($CurrentNode$) \& $\notin$ buffer(}\emph{\scriptsize{}$Node_{i}$}{\scriptsize{})
}\textbf{\scriptsize{}do\vspace{-0.03 cm}}{\scriptsize \par}

{\scriptsize{}\hspace{0.6 cm}}\textbf{\scriptsize{}if}{\scriptsize{}
(}\emph{\scriptsize{}$Node_{i}$.}{\scriptsize{}getWeightTo($Destination_{j}$)
$>$ }{\scriptsize \par}

{\scriptsize{}\hspace{3 cm}$CurrentNode$}\emph{\scriptsize{}.}{\scriptsize{}getWeightTo($Destination_{j}$))}\textbf{\scriptsize{}\vspace{-0.03 cm}}{\scriptsize \par}

{\scriptsize{}\hspace{0.9 cm}}\textbf{\scriptsize{}then}{\scriptsize{}
$CurrentNode$.replicateTo(}\emph{\scriptsize{}$Node_{i}$}{\scriptsize{},
$Message_{j}$)}\textbf{\scriptsize{}\vspace{-0.03 cm}}{\scriptsize \par}

{\scriptsize{}\hspace{0.6 cm}}\textbf{\scriptsize{}else}{\scriptsize \par}

{\scriptsize{}\hspace{0.9 cm}receive(}\emph{\scriptsize{}$Node_{i}$.}{\scriptsize{}Importance)}\textbf{\scriptsize{}\vspace{-0.03 cm}}{\scriptsize \par}

{\scriptsize{}\hspace{0.9 cm}}\textbf{\scriptsize{}if}{\scriptsize{}
(}\emph{\scriptsize{}$Node_{i}$.}{\scriptsize{}importance $>$ $CurrentNode$}\emph{\scriptsize{}.}{\scriptsize{}importance)}\textbf{\scriptsize{}\vspace{-0.03 cm}}{\scriptsize \par}

{\scriptsize{}\hspace{0.9 cm}}\textbf{\scriptsize{}then}{\scriptsize{}
$CurrentNode$.replicateTo(}\emph{\scriptsize{}$Node_{i}$}{\scriptsize{},
$Message_{j}$)}\textbf{\scriptsize{}\vspace{-0.1 cm}}{\scriptsize \par}

\textbf{\scriptsize{}end}
\end{algorithm}
\vspace{-0.3 cm}

As mentioned before, \emph{dLifeComm} combines the notion of community,
as \emph{Bubble} \emph{Rap}, and social strength for forwarding: when
a user has a message to another user in a different community, it
forwards the message towards the destination's community using \emph{TECDi}.
The assumption is that users with higher importance have higher probability
to reach the destination's community faster. When the destination's
community is reached, forwarding is done towards the destination,
by replicating the message to users with higher social strength (\emph{TECD})
towards the destination, and not higher centrality, as in \emph{Bubble}
\emph{Rap}. The main goal is to avoid congestion points.

\emph{dLifeComm's} operation is rather simple (cf. Alg. \ref{alg:DlifeComm}):
when the $CurrentNode$ meets a \emph{$Node_{i}$}, it gets a list
of all neighbors of \emph{$Node_{i}$} and its weights towards them
(\emph{\scriptsize{}$Node_{i}$}{\scriptsize{}.WeightsToAllneighbors}
computed based on Eq. \ref{eq:3}) in the current daily sample $\Delta T_{k}$.
If \emph{$Node_{i}$ }belongs to the same community as the destination
of $Message_{j}$, the message is replicated if the weight of \emph{$Node_{i}$}
towards the destination is greater than $CurrentNode$'s weight towards
this destination. If \emph{$Node_{i}$ }belongs to a different community,
$CurrentNode$ receives \emph{$Node_{i}$}'s importance, and messages
are replicated if \emph{$Node_{i}$}'s importance is greater than
that of the $CurrentNode$. 

As \emph{Bubble Rap}, \emph{dLifeComm} allows users - not in the destination
community - to delete messages already delivered to such community,
to avoid useless replications. It is worth noting that \emph{dLifeComm}'s
algorithm is different from that of \emph{Bubble} \emph{Rap} as it
uses \emph{TECD}/\emph{TECDi} instead of local/global centralities
for forwarding within/between communities.\vspace{-0.2 cm}

\begin{algorithm}[H]
\protect\caption{\label{alg:DlifeComm}Forwarding with \emph{dLifeComm}}

\textbf{\scriptsize{}begin}{\scriptsize \par}

{\scriptsize{}\hspace{0.1 cm}}\textbf{\scriptsize{}foreach}\emph{\scriptsize{}
$Node_{i}$ }{\scriptsize{}encountered by $CurrentNode$ }\textbf{\scriptsize{}do}{\scriptsize \par}

{\scriptsize{}\hspace{0.3 cm}receive(}\emph{\scriptsize{}$Node_{i}$}{\scriptsize{}.WeightsToAllneighbors)}\textbf{\scriptsize{}\vspace{-0.01 cm}}{\scriptsize \par}

{\scriptsize{}\hspace{0.3 cm}}\textbf{\scriptsize{}foreach}{\scriptsize{}
$Message_{j}$}\textbf{\scriptsize{} }{\scriptsize{}$\in$}\textbf{\scriptsize{}
}{\scriptsize{}buffer.($CurrentNode$) \& $\notin$ buffer(}\emph{\scriptsize{}$Node_{i}$}{\scriptsize{})
}\textbf{\scriptsize{}do\vspace{-0.03 cm}}{\scriptsize \par}

{\scriptsize{}\hspace{0.6 cm}}\textbf{\scriptsize{}if}{\scriptsize{}
(}\emph{\scriptsize{}$Node_{i}$}{\scriptsize{}.isInCommunityOf($Destination_{j}$))}\textbf{\scriptsize{}\vspace{-0.03 cm}}{\scriptsize \par}

{\scriptsize{}\hspace{0.9 cm}}\textbf{\scriptsize{}if}{\scriptsize{}
(}\emph{\scriptsize{}$Node_{i}$.}{\scriptsize{}getWeightTo($Destination_{j}$)
$>$ }{\scriptsize \par}

{\scriptsize{}\hspace{3 cm}$CurrentNode$}\emph{\scriptsize{}.}{\scriptsize{}getWeightTo($Destination_{j}$))}\textbf{\scriptsize{}\vspace{-0.03 cm}}{\scriptsize \par}

{\scriptsize{}\hspace{0.9 cm}}\textbf{\scriptsize{}then}{\scriptsize{}
$CurrentNode$.replicateTo(}\emph{\scriptsize{}$Node_{i}$}{\scriptsize{},
$Message_{j}$)}\textbf{\scriptsize{}\vspace{-0.03 cm}}{\scriptsize \par}

{\scriptsize{}\hspace{0.6 cm}}\textbf{\scriptsize{}else}{\scriptsize \par}

{\scriptsize{}\hspace{0.9 cm}receive(}\emph{\scriptsize{}$Node_{i}$.}{\scriptsize{}Importance)}\textbf{\scriptsize{}\vspace{-0.03 cm}}{\scriptsize \par}

{\scriptsize{}\hspace{0.9 cm}}\textbf{\scriptsize{}if}{\scriptsize{}
(}\emph{\scriptsize{}$Node_{i}$.}{\scriptsize{}importance $>$ $CurrentNode$}\emph{\scriptsize{}.}{\scriptsize{}importance)}\textbf{\scriptsize{}\vspace{-0.03 cm}}{\scriptsize \par}

{\scriptsize{}\hspace{0.9 cm}}\textbf{\scriptsize{}then}{\scriptsize{}
$CurrentNode$.replicateTo(}\emph{\scriptsize{}$Node_{i}$}{\scriptsize{},
$Message_{j}$)}\textbf{\scriptsize{}\vspace{-0.1 cm}}{\scriptsize \par}

\textbf{\scriptsize{}end}
\end{algorithm}
\vspace{-0.3 cm}

\section{\emph{dLife} Evaluation\label{sec:Evaluation-and-Results}}

This section starts by describing the evaluation methodology and experimental
settings. Then, our considerations about the results obtained when
comparing \emph{dLife} with \emph{dLifeComm} and \emph{Bubble} \emph{Rap}
are presented considering two scenarios: one based on simulations
built with different mobility patterns, and another based on real
human traces.

\subsection{Evaluation Methodology}

Performance analysis of \emph{dLife}, \emph{dLifeComm}, and \emph{Bubble
Rap} is done on the Opportunistic Network Environment (ONE). 

Each simulation, in both scenarios, is run ten times (with different
random number generator seeds) to provide results with a 95\% confidence
interval. All results are analyzed in terms of average delivery probability
(i.e., ratio between the number of delivered messages and total number
of created messages), average cost (i.e., number of replicas per delivered
message), and average latency (i.e., time elapsed between message
creation and delivery).

\subsection{Experimental Settings \label{sub:Settings}}

The heterogeneous simulation scenario is part of the Helsinki city
and has 150 nodes distributed in 8 groups of people and 9 groups of
vehicles. Nodes are equipped with a WiFi interface (11 Mbps/100 m).
One vehicle group represents police patrols and follows the \emph{Shortest
Path Map Based Movement} where nodes randomly choose a destination
point and take the shortest path to it. Their waiting times are between
100 and 300 seconds. The remaining groups represent buses, each composed
of 2 vehicles following the \emph{Bus Movement} and with waiting times
between 10 and 30 seconds. Vehicles speeds range from 7 to 10 m/s.

People follow the \emph{Working Day Movement} with walking speeds
ranging from 0.8 to 1.4 m/s, but can also use buses. Each group has
different meeting spots, offices, and home locations. People spend
8 hours at work and present 50\% probability of having an evening
activity after leaving work. In the office, nodes move around and
have pause times ranging from 1 minute to 4 hours. Evening activities
can be done alone or in a group, and can last between 1 and 2 hours. 

For the experiments based on real human traces, we use the Cambridge
traces \cite{cambridge-haggle-imote-content-2006-09-15} between 36
nodes. Data was collected in different locations for two months while
Cambridge University students moved performing their daily activities.

Traffic load comes from a file previously generated with established
source/destination pairs, where a total of 6000 messages are generated
in each scenario.

Message TTL values are set at 1, 2 and 4 days, as well as 1 and 3
weeks. Since we want a fair comparison against \emph{Bubble Rap},
we choose the TTL values in which \emph{Bubble Rap} has the best performance
behavior in terms of delivery probability and cost \cite{bubble2011},
as well as TTL values that can represent the different applications
that cope with opportunistic networks. Message size ranges from 1
to 100 kB. The buffer space is of 2 MB to create a realistic scenario,
as users may not be willing to share all of the storage capacity of
their devices. Message and buffer size comply with the universal evaluation
framework that we have proposed \cite{latincom} based on the evidence
that prior-art on opportunistic routing follows completely different
evaluation settings, making the assessment a challenging task.

To assess the performance of \emph{dLifeComm }and \emph{Bubble Rap},
we consider K-Clique and cumulative window algorithms for community
formation and node centrality computation as proposed by Hui et al.
\cite{bubble2011}. The parameter $k$ ($=5$) is chosen based on
simulations in which \emph{Bubble Rap} has the best overall performance
in terms of the considered evaluation metrics.

As \emph{dLife} considers daily samples (cf. Section \ref{sec:Proposal}),
our findings (omitted due to limited space) show that 24 daily samples
brings \emph{dLife} to its best. The reason is that the shorter the
samples (i.e., one hour), the more refined the information on social
strength and users' importance is.

\subsection{Experimental Results}

We start by providing some considerations about our findings: the
average number of contacts per hour is of approximately 920 in the
heterogeneous scenario and of approximately 29 in the human traces.
Moreover, contacts are more sporadic in the trace scenario than in
the heterogeneous one, in which contact frequency is more homogeneous.
We also notice that the average number of unique communities is higher
in the heterogeneous scenario (\textasciitilde{}69) than in the trace
scenario (\textasciitilde{}6.7). Furthermore, most of the created
communities encompasses all the existing nodes (150 for the heterogeneous
simulations, and 36 for trace). This means that independently of the
level of contact homogeneity, nodes are still well connected.\vspace{-0.2 cm}

\begin{figure}[H]
\subfigure[Heterogeneous scenario]{\label{fig:adp_scenario}

\includegraphics[scale=0.55]{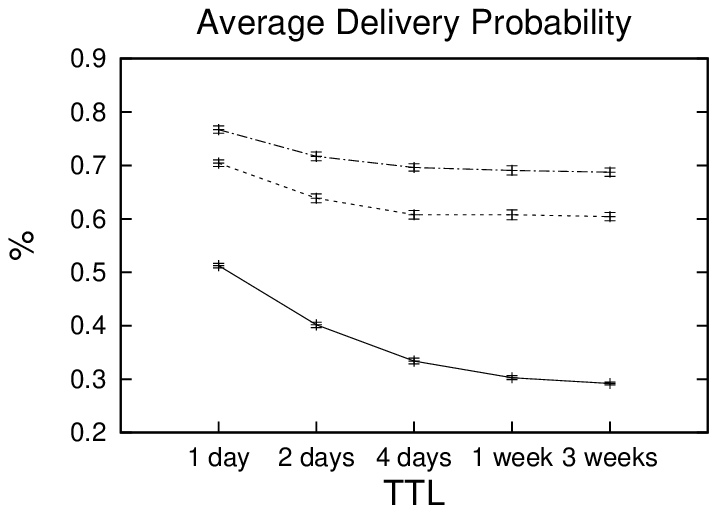}}\hspace{-0.09cm} \subfigure[Cambridge traces]{
\label{fig:adp_traces} \includegraphics[scale=0.55]{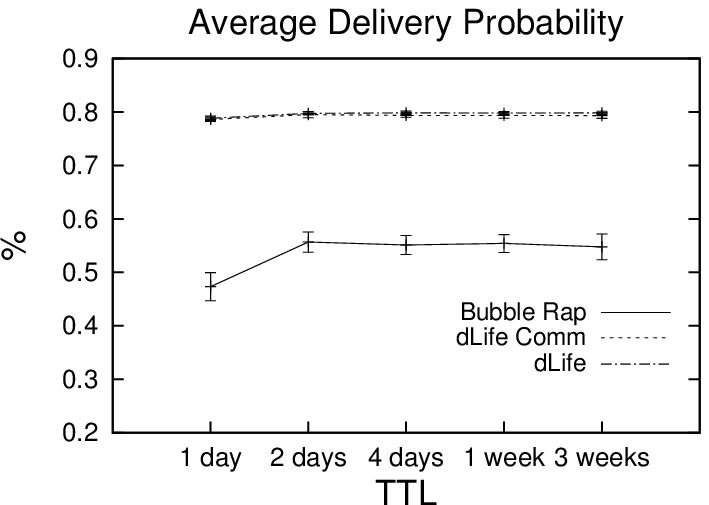} }
\label{fig:adp} \vspace{-0.4cm}
 \protect\caption{Average delivery probability}
\vspace{-0.4 cm}
\end{figure}

Figs. \ref{fig:adp_scenario} and \ref{fig:adp_traces} show the average
delivery probability: for the simulated heterogeneous scenario (cf.
Fig. \ref{fig:adp_scenario}), \emph{dLife} and \emph{dLifeComm} have
a performance 39.5 and 31.2 percentage points better than \emph{Bubble}
\emph{Rap}, respectively. The main reason for that is that \emph{Bubble}
\emph{Rap} has to go through the process of forming communities to
perform suitable forwarding. Since communities are not immediately
available, \emph{Bubble Rap} relies on node global centrality to increase
the probability of reaching destinations. However, in this scenario,
the centrality of nodes is very heterogeneous where a few nodes (\textasciitilde{}17\%)
have very high centrality and the remaining nodes have mid/low centrality.
Since most of the messages are originated in nodes with mid/low centrality,
this results in a increase in message replication as \emph{Bubble}
\emph{Rap} replicates when meeting a node with higher centrality.
Such behavior quickly exhausts buffer space, which worsens as TTL
increases since messages are allowed to live longer in the system,
having higher probability to be replicated. Both versions of \emph{dLife}
also experience buffer exhaustion as TTL increases, and \emph{dLifeComm
}is affected by the community formation. However, since replication
occurs wisely due to \emph{dLife}'s capability of capturing the dynamism
of nodes' behavior, these effects are mitigated.

With real traces (cf. Fig. \ref{fig:adp_traces}), \emph{dLife} and
\emph{dLifeComm} still have better performance (reaching up to 31.5
and 31.3 percentage points, respectively) than \emph{Bubble} \emph{Rap},
which shows a similar behavior as reported by Hui et al. \cite{bubble2011},
where delivery probability increases with TTL, since K-Clique creates
an average of 6.7 communities encompassing almost all nodes in each
one. In this situation a 2-day TTL is enough to reach a node in the
destination community increasing the probability of delivery. However,
since \emph{Bubble} \emph{Rap} relies on a central local node to deliver
inside a community, and since there is still a probability that such
node may not be well connected with the destination, the probability
of delivery does not benefit from a higher TTL.

The good performance of both versions of \emph{dLife} is due to network
dynamics (from users' daily routines). This allied to the network
structure (i.e., communities), made \emph{dLifeComm} outperform \emph{Bubble}
\emph{Rap}, but still suffering with the community formation overhead;
while by only considering such dynamics, \emph{dLife} turns out to
be the best proposal. We believe that the similar performance behavior
of both proposals in the human trace scenario is due to the fact that
very little communities are formed and most of the nodes belong to
such communities, thus reducing the effect of the overhead seen in
the heterogeneous scenario. Additionally, results suggest that the
usage of centrality has a higher impact (i.e., negative) than the
usage of community formation, as centrality creates bottlenecks: this
can be easily seen when comparing \emph{dLifeComm} (which combines
the notion of community and \emph{TECD}/\emph{TECDi}) and \emph{Bubble
Rap }(which combines the notion of community and centrality).\vspace{-0.2 cm}

\begin{figure}[H]
\subfigure[Heterogeneous scenario]{\label{fig:ac_scenario}

\includegraphics[scale=0.55]{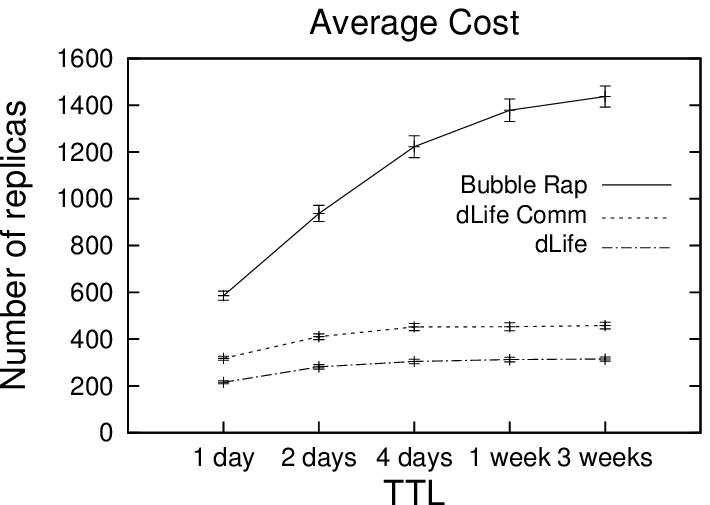}}\subfigure[Cambridge traces]{\label{fig:ac_traces}
\includegraphics[scale=0.55]{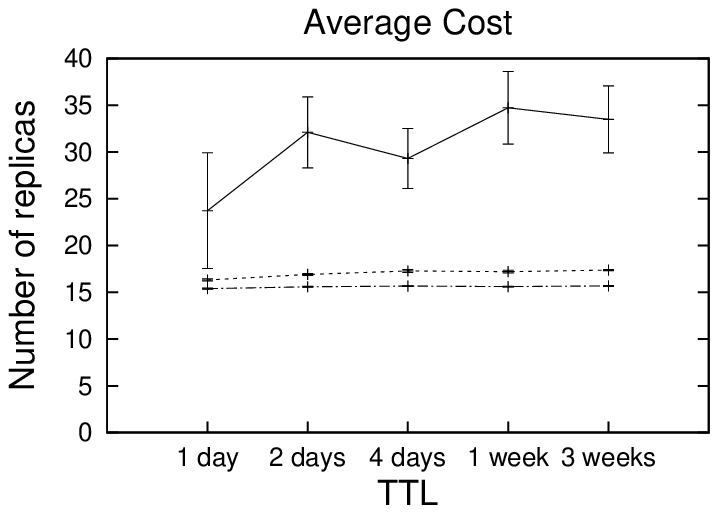} } \label{fig:ac} \vspace{-0.4cm}
 \protect\caption{Average cost}
\vspace{-0.4 cm}
\end{figure}

Next we look at the average cost (cf. Figs. \ref{fig:ac_scenario}
and \ref{fig:ac_traces}). We observe in the simulated heterogeneous
scenario (cf. Fig. \ref{fig:ac_scenario}) that \emph{dLife} and \emph{dLifeComm}
are cost effective. They produce up to 78\% and 68\% less replicas
than \emph{Bubble} \emph{Rap}, respectively. This good behavior reflects
the wise forwarding decisions that both proposals are able to perform
(based on \emph{TECD} and \emph{TECDi}). It is indeed an indication
that \emph{dLife} is able to derive a clearer social graph, based
on edges with high social strength and vertices with higher importance.
Regarding \emph{Bubble Rap}, its cost is expected to increase with
TTL: despite getting rid of a message when it reaches the destination's
community, to avoid further replication, other replicas continue to
be made by other carriers, which explains \emph{Bubble Rap}'s higher
cost.

The real trace scenario (cf. Fig. \ref{fig:ac_traces}) still shows
the lower cost of \emph{dLife} and \emph{dLifeComm} in relation to
\emph{Bubble Rap} (reaching up to 55\% and 50.5\% less, respectively).
The cost reduction for \emph{Bubble} \emph{Rap} with a 4-day TTL is
due to the sporadic nature of contacts in this scenario. This results
in a lower average number of replicas created as there are only few
nodes to receive such copies at the time of exchange.

\begin{figure}[H]
\subfigure[Heterogeneous scenario]{\label{fig:al_scenario}

\includegraphics[scale=0.55]{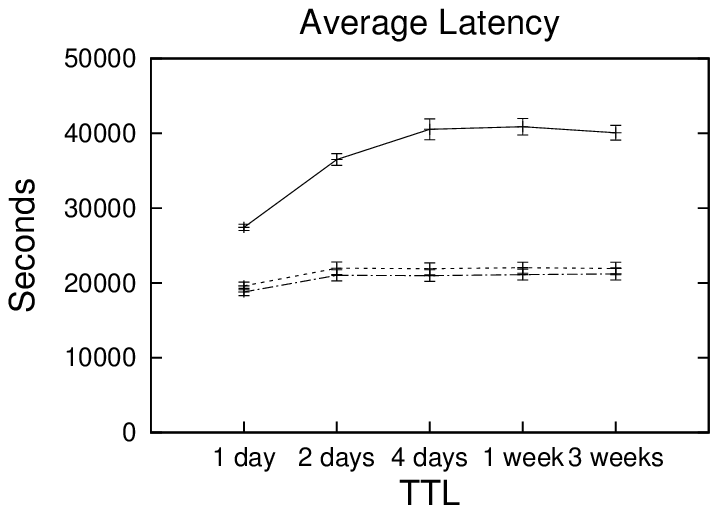}}\subfigure[Cambridge traces]{
\label{fig:al_traces} \includegraphics[scale=0.55]{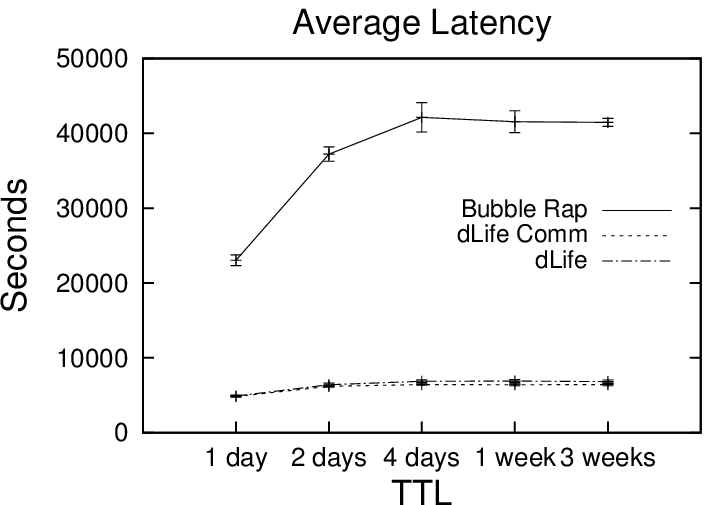} }
\label{fig:al} \vspace{-0.4cm}
 \protect\caption{Average latency}
\vspace{-0.4 cm}
\end{figure}

Regarding the average latency (cf. Figs. \ref{fig:al_scenario} and
\ref{fig:al_traces}), in the heterogeneous scenario (cf. Fig. \ref{fig:al_scenario}),
\emph{dLife} and \emph{dLifeComm} deliver messages with lower latency
(48.3\% and 46.1\% less, respectively) than \emph{Bubble} \emph{Rap}.
It is easily observed the advantage of taking wiser decisions by using
\emph{dLife}: messages are only forwarded in the presence of strong
social links or highly important nodes in the current daily sample.
Thus, by considering stronger social links with the destination or
more important encountered nodes in specific daily samples, messages
are delivered faster, since the probability of them coming in contact
with the destination in the near future is higher. \emph{Bubble Rap}
does not capture such dynamism which leads it to create replicas that
take more time to reach the destination due to the weaker social ties
of the carrier with the destination. These results suggest that considering
the dynamism of daily routines allows nodes to select the best forwarder
in different daily samples, while centrality leads to the identification
of a node that may be well connected in average for the complete duration
of the experiment, but not in all daily samples.\vspace{-0.2 cm}

In the real trace scenario (cf. Fig. \ref{fig:al_traces}), \emph{dLife}
and \emph{dLifeComm} deliver messages faster (83.7\% and 84.7\% less,
respectively) than \emph{Bubble} \emph{Rap} in comparison with the
heterogeneous scenario. Despite the sporadic contacts of the real
trace scenario, some nodes are still well connected and, since both
versions of \emph{dLife} are able to identify the best social connections
(i.e., independently of the notion of community), messages are replicated
to nodes which have the highest probability to meet destinations in
the near future, decreasing the distance (i.e., hops) to reach the
destination, which in turn reduces the delivery time. \emph{Bubble}
\emph{Rap} also experiences a reduction in the distance to reach destinations,
but its latency behavior remains almost the same in both scenarios.
We believe this is due to the fact that most of the existing nodes
in the studied scenarios belong to the formed communities, as earlier
noted. Since these communities are almost the same for the duration
of the experiments, \emph{Bubble} \emph{Rap} relies solely on node
centrality, which does not capture the dynamism of users' behavior,
and thus messages need more time to reach destinations. Despite of
considering community formation, \emph{dLifeComm} is less affected
since it also considers the node importance to propagate messages.

\section{Conclusions and Future Work\label{sec:Conclusions}}

Since social information is quite useful to aid data forwarding in
opportunistic networks, we introduce \emph{dLife,} which combines
the \emph{TECD} and \emph{TECDi} utility functions to derive, from
users' social daily routines, the social strength among users and
their importance. Our findings show that by incorporating the dynamism
of users' social daily behavior in opportunistic routing wiser forwarding
decisions are performed, leading to better delivery probability, cost
and latency than proposals based only on social structures, i.e.,
\emph{Bubble} \emph{Rap}. Moreover, by comparing \emph{Bubble} \emph{Rap}
with \emph{dLifeComm}, a solution that combines network structure
(i.e., communities) with network dynamics (i.e., daily routines),
we show that the usage of centrality has a higher impact (i.e., negative)
in the system performance than the usage of detected communities.

As future work, we plan to improve \emph{dLife}'s performance with
the introduction of randomness: it has been shown that even with complete
knowledge on the social relationship among users, delivery probability
does not reach its maximum \cite{people}. Additionally, we will extend
\emph{dLife} to have a point-to-multipoint behavior and test it with
real traces encompassing large number of nodes (e.g., MIT Reality
mining project).

\section*{Acknowledgment}

\begin{singlespace}
\noindent To FCT for supporting UCR (PTDC/EEA-TEL/103637/2008) project
and Mr. Moreira's PhD grant (SFRH/BD/62761/2009).{\large{}\bibliographystyle{ieeetr}
\bibliography{bib-or}
}\end{singlespace}

\end{document}